\documentclass[twocolumn,nofootinbib,amsmath,prd,aps,superscriptaddress,
tightenlines,preprintnumbers,nobalancelastpage]{revtex4}
\usepackage{amsmath, amsfonts, amsthm, amssymb}
\allowdisplaybreaks
\usepackage{amsmath}
\usepackage{units}
\usepackage{subcaption}
\usepackage{ifpdf}
\usepackage{accents}
\usepackage{tikz}
\usetikzlibrary{decorations.pathmorphing}
\usetikzlibrary{decorations.markings}
\usepackage{tikz-feynman}
\tikzfeynmanset{compat=1.0.0}
\usepackage{slashed}
\usepackage{bbold}
\usepackage{graphicx} 
\usepackage{cancel}
\usepackage{dsfont}
\usepackage{physics}
\usepackage{ytableau}
\usepackage{braket}
\usepackage{booktabs}
\usepackage{multirow}
\usepackage{booktabs}
\numberwithin{equation}{section}

\usepackage{catchfile}  
\newcommand{\getenv}[2][]{%
  \CatchFileEdef{\temp}{"|kpsewhich --var-value #2"}{}%
  \if\relax\detokenize{#1}\relax\temp\else\let#1\temp\fi
}
\getenv[\USER]{USER}

\RequirePackage[colorlinks=true,urlcolor=blue,anchorcolor=blue,citecolor=blue,filecolor=blue,
               linkcolor=blue,menucolor=blue,linktocpage=true,pdfproducer=medialab,pagebackref=true]{hyperref}
\hypersetup{
  colorlinks   = true, 
  urlcolor     = mygreen, 
  linkcolor    = blue, 
  citecolor   = blue 
}
 \usepackage[capitalise]{cleveref}
\usepackage{csquotes}

\newcommand{\nn}{\nonumber}

\definecolor{mygreen}{rgb}{0.0, 0.5, 0.0}

\usepackage[capitalise]{cleveref}
\usepackage{subcaption}

\def\abs#1{\left| #1\right|}

\usepackage{orcidlink}

\usepackage[font=small,labelfont=bf,justification=centerlast]{caption}
\usepackage{mathtools}

\begin{document}

\title{
A High-Quality Axion from Exact SUSY Chiral Dynamics
}

\author{Tony Gherghetta \orcidlink{0000-0002-8489-1116}}
\email{tgher@umn.edu}
\affiliation{School of Physics and Astronomy, University of Minnesota, Minneapolis, Minnesota 55455, USA}

\author{Hitoshi Murayama \orcidlink{0000-0001-5769-9471}}
\email{hitoshi@berkeley.edu, hitoshi.murayama@ipmu.jp, Hamamatsu Professor}
\affiliation{The Leinweber Institute for Theoretical Physics, University of California, Berkeley, California 94720, USA}
\affiliation{Kavli Institute for the Physics and Mathematics of the
  Universe (WPI), University of Tokyo,
  Kashiwa 277-8583, Japan}
\affiliation{Ernest Orlando Lawrence Berkeley National Laboratory, Berkeley, California 94720, USA}

\author{Bea Noether\,\orcidlink{0000-0002-2947-3210}}
\email{bea\_noether@berkeley.edu}
\affiliation{The Leinweber Institute for Theoretical Physics, University of California, Berkeley, California 94720, USA}
\affiliation{Ernest Orlando Lawrence Berkeley National Laboratory, Berkeley, California 94720, USA}

\author{Pablo Qu\'ilez\orcidlink{0000-0002-4327-2706}}
\email{pquilez@ucsd.edu}
\affiliation{\it Department of Physics University of California, San Diego, California, USA
}

\begin{abstract}
We use supersymmetric chiral dynamics perturbed by anomaly-mediated supersymmetry breaking to obtain a high-quality, composite axion that solves the strong CP problem. The strong dynamics arises from a supersymmetric SU(10) chiral gauge theory with massless matter chiral superfields.
This leads to a stable, nonsupersymmetric vacuum, calculated exactly, where a spontaneously broken $U(1)$ global symmetry, identified with the Peccei-Quinn symmetry, gives rise to a composite QCD axion. The chiral gauge theory also admits a discrete  $\mathbb{Z}_{4}$ (or $\mathbb{Z}_{8,16}$) gauge symmetry that forbids PQ-violating operators up to dimension eight.
An extension to an $SU(14)\times \mathbb{Z}_{12}$ chiral gauge theory incorporates $SO(10)$ grand unification where the unification scale is identified with the PQ-breaking scale and PQ-violating operators are forbidden up to dimension 20.
The supersymmetry breaking scale, near 100 TeV, ameliorates the Higgs hierarchy problem,  while colored NGBs may be detected at future colliders via decays to gluons or form heavy isotopes.

\end{abstract}
\maketitle



\section{Introduction}

Supersymmetric chiral dynamics provides a novel framework to explore strongly-coupled gauge theories, where confinement and anomaly matching naturally generate composite states and accidental global symmetries. These features are especially useful for building models of the QCD axion, which solve the strong CP problem. In particular, the strong dynamics can spontaneously break the Peccei-Quinn (PQ) symmetry~\cite{Peccei:1977hh}, with the axion decay constant scale naturally tied to the dynamical scale and where the resulting pseudo Nambu-Goldstone boson (NGB) or axion~\cite{Weinberg:1977ma,Wilczek:1977pj} is identified as a composite state. In addition, the strong dynamics can also allow the PQ symmetry to be of sufficiently high quality that is robust against quantum-gravity effects~\cite{Georgi:1981pu,Kamionkowski:1992mf,Holman:1992us,Kallosh:1995hi,Barr:1992qq,Ghigna:1992iv,Alonso:2017avz,Alvey:2020nyh}.  At the same time, supersymmetry addresses the Higgs hierarchy problem by stabilizing the electroweak scale (see, for example, \cite{Martin:1997ns}), making it a natural framework to pursue solutions to both the strong CP and hierarchy problems. 

Early composite axion models largely relied on nonsupersymmetric dynamics~\cite{Kim:1984pt,Choi:1985cb} to generate axions with phenomenologically viable axion decay constants, but did not attempt to address the axion quality problem. Subsequent work~\cite{Randall:1992ut,Harigaya:2015soa,Redi:2016esr,Lillard:2018fdt,Gherghetta:2025fip} explored the use of strong dynamics including nonsupersymmetric and supersymmetric examples to improve the robustness of the PQ symmetry. More recently, theories with a simple confining chiral group have been proposed \cite{Gavela:2018paw,Ardu:2020qmo,Contino:2021ayn} including models that are consistent with a dual holographic description~\cite{Cox:2019rro,Cox:2023dou}.

One of the key assumptions of all chiral models is that the chiral symmetries (and in particular the PQ symmetry) are spontaneously broken below the confining scale. This is fundamental for the viability of the model since the composite axion is the pseudo-NGB of the spontaneously broken PQ symmetry. Indeed, confinement without chiral symmetry breaking would lead to extra massless colored fermions, which are ruled out. Describing the IR degrees of freedom of confining theories is one of the open problems in theoretical physics. Only a few exact results are known, particularly for vector-like theories, along with certain consistency conditions such as 't Hooft anomaly matching \cite{tHooft:1979rat}. However, recent developments \cite{Murayama:2021xfj,Csaki:2021xhi,Csaki:2021aqv} make use of exact results in supersymmetric gauge theories to infer the low energy states in both supersymmetric and non-supersymmetric confining theories. These techniques can be applied to elucidate which composite axion models based on chiral gauge groups give rise to viable solutions to the strong CP problem.

In this work, we consider a supersymmetric $SU(10)$ chiral gauge theory with massless matter chiral supermultiplets
where the QCD gauge group weakly gauges an unbroken $Sp(6)$ global symmetry. As shown in Ref~\cite{Csaki:2021xhi}, this theory can be perturbed by anomaly-mediated supersymmetry breaking (AMSB) which gives rise to a stable, nonsupersymmetric vacuum with spontaneously broken global symmetries. Interestingly, an anomaly-free $U(1)$ can be identified with the PQ symmetry that is spontaneously broken leading to a composite axion. The PQ symmetry is, however, anomalous with respect to QCD, and therefore the composite axion receives the usual axion mass contribution from nonperturbative QCD.

Moreover, we identify a discrete global symmetry $\Gamma={\mathbb Z}_{16}$, in order to forbid Planck suppressed PQ-violating operators that are responsible for misaligning the axion potential minimum. In general, global symmetries are broken by quantum gravitational effects and therefore must be ``gauged"~\cite{Krauss:1988zc,Preskill:1990bm} (i.e. embedded into a continuous local symmetry) in order to be preserved under gravity. We show that the $\Gamma={\mathbb Z}_{16}$ discrete symmetry is indeed compatible with both $SU(10)$ and $SU(3)_c$ instantons as well as gravity and therefore since it is anomaly free, can be gauged. This anomaly-free discrete global symmetry then forbids Planck suppressed PQ-violating operators to very high order, thereby robustly addressing the axion-quality problem in our supersymmetric model. In addition, the discrete subgroups ${\mathbb Z}_{4}({\mathbb Z}_{8})$ 
can also be imposed that forbid PQ-violating operators provided the axion-photon coupling satisfies $g_{a\gamma}^{-1} \lesssim 10^{13}(10^{17})$~GeV. Discrete gauge symmetries were also used in \cite{Babu:2002ic,Lee:2011dya,Harigaya:2013vja,Bhattiprolu:2021rrj} to suppress PQ-violating operators in supersymmetric models but without invoking any strong dynamics.

Finally, the $SU(10)$ model can be extended to an $SU(14)$ model that allows for grand unification. In particular, when the grand unified group $SO(10)$ weakly gauges the $SU(10)$ global symmetry of the $SU(14)$ chiral theory, the grand unification scale can identified with the axion decay constant scale $f_a$. This theory also admits a discrete ${\mathbb Z}_{12}$ gauge symmetry which forbids PQ-violating operators up to dimension 20, that can be further enhanced to ${\mathbb Z}_{24}$.

\section{The nonsupersymmetric $SU(5)$ model}
\label{sec:GIQYmodel}

We first review the model in Ref.~\cite{Gavela:2018paw} that considered a nonsupersymmetric, $SU(5)$ chiral gauge theory to identify the QCD axion as a composite NGB. The Peccei-Quinn symmetry was identified as a global symmetry with a Planck-scale violation that was accidentally suppressed by six-fermion (dimension 9) terms.

The $SU(5)$ gauge anomalies are cancelled with $n$ fermion pairs, $\bar{F}^i(\mathbf{\bar 5})$ and $A_j({\bf 10})$, $(i,j=1, \cdots, n)$ where the representations under the gauge and global symmetries are shown in \cref{tab:SU5}. Both global symmetries, $SU(n)_A$ and $SU(n)_{\bar{F}}$, are assumed to contain an $n$-dimensional real representation ${\bf R}$ of QCD $SU(3)_c$  which may be irreducible ({\it e.g.}\/, {\bf 8}) or reducible ({\it e.g.}\/, ${\bf 3} + \mathbf{\bar 3}$). The non-Abelian global symmetry $SU(n)_A \times SU(n)_{\bar{F}}$, must be spontaneously broken since the t'Hooft anomalies cannot be matched. However, the $U(1)_{\rm PQ}$ symmetry was assumed to be spontaneously broken, while preserving QCD $SU(3)_c$, even though the corresponding t'Hooft anomalies are known to potentially be matched.

\begin{table*}[htb]
\begin{center}
\ytableausetup{boxsize=0.7em}
\ytableausetup{aligntableaux=center}
\hspace*{-8pt}\begin{tabular}{|c|c||c|c|c|c||c|c|} \hline
& $SU(5)$  &  $SU(n)_A$ & $SU(n)_{\bar{F}}$ & $U(1)_{\rm PQ}$
& $\mathbb{Z}_{3n}$ & $SO(n)_{\rm diag}$ & $SU(3)_c$\\\hline \hline
$A$ & \tiny\ydiagram[]{1,1}  &$\overline{\tiny\ydiagram[]{1}}$ & {\bf 1} & $-1$ 
& {\bf 1} & \tiny\ydiagram[]{1} & {\bf R}\\[2.5pt] \hline
$\bar{F}$ & $\overline{\tiny\ydiagram[]{1}}$  & {\bf 1} & \tiny\ydiagram[]{1} & $+3$ 
& {\bf 0} & \tiny\ydiagram[]{1} & {\bf R}\\ \hline \hline
$A\bar{F}\bar{F}$ & {\bf 1}  & $\overline{\tiny\ydiagram[]{1}}$ & $\tiny\ydiagram[]{2}$ & $+5$ 
& {\bf 1} & $\tiny\ydiagram[]{1}$ & {\bf R}+\dots\\ \hline
$AAA\bar{F}$ & {\bf 1} 
 & $\overline{\tiny\ydiagram[]{2}} \times \overline{\tiny\ydiagram[]{1}}$
& \tiny\ydiagram[]{1} & $0$ 
& {\bf 3} & ${\bf 1}$ & {\bf 1}+\dots\\ \hline
\end{tabular}
\end{center}
\caption{The representations of the fermion matter content in the $SU(5)$ model~\cite{Gavela:2018paw}, including the composite fermion $A\bar{F}\bar{F}$ and boson $AAA\bar{F}$. 
}
\label{tab:SU5}
\end{table*}

In order to verify whether these global symmetry breaking assumptions are justified, we propose the following dynamics which satisfies the 't Hooft anomaly matching conditions in a highly non-trivial fashion.
We postulate that the $SU(n)_A \times SU(n)_{\bar{F}}$ global symmetry is broken to the $SO(n)_{\rm diag}$ subgroup, while $U(1)_{\rm PQ}$ remains unbroken. This is achieved by the order parameter
\begin{align}
    \epsilon_{abcde} \langle (A_i^{ab} A_j^{cd}) (A_k^{ef} \bar{F}^l_f) \rangle
    \propto \delta^{ij} \delta_k^l \neq 0\,,
    \label{eq:orderp}
\end{align}
where $a,\dots ,f$ are SU(5) indices and $i,j,k,l$ are flavor indices.
The spinor indices are not shown but are contracted within each bracket. Note that there are no fermion bilinear gauge-invariant operators in the theory because it is chiral and thus the PQ symmetry arises accidentally. 
The lowest dimension possible order operator for symmetry breaking is quartic and the only such operator is given in \eqref{eq:orderp}. 
However, the four-fermion operator in \eqref{eq:orderp} has zero PQ charge, and therefore it is postulated that $\delta_k^l$ only breaks the $SU(n)_A \times SU(n)_{\bar{F}}$ flavor symmetry down to the diagonal subgroup $SU(n)_{\rm diag}$, and $\delta^{ij}$ further breaks $SU(n)_{\rm diag}$ to its $SO(n)_{\rm diag}$ subgroup. In addition, we postulate that there are massless composite fermions
\begin{align}
    A^{ab}_i(\bar{F}^i_a \bar{F}^j_b), \label{eq:AFF}
\end{align} 
that transform as a vector under $SO(n)_{\rm diag}$ with $U(1)_{\rm PQ}$ charge $+5$. 

Our postulate satisfies the following 't Hooft anomaly matching conditions 
\begin{align}
    &U(1)_{\rm PQ}\times [SO(n)_{\rm diag}]^2 \\ 
    & \qquad {\rm UV:}\ 10\cdot(-1)+5\cdot(+3)=5 \nonumber \\
    & \qquad {\rm IR:}\ 5\cdot(+1)=5 \nonumber\\
    &[U(1)_{\rm PQ}]^3 \\ 
    & \qquad {\rm UV:}\ 10n\cdot(-1)^3+5n\cdot(+3)^3=125n \nonumber \\
    & \qquad {\rm IR:}\ n\cdot(+5)^3=125n \nonumber\\
    &U(1)_{\rm PQ}\times [{\rm (gravity)}]^2 \\ 
    & \qquad {\rm UV:}\ 10n\cdot(-1)+5n\cdot(+3)=5n \nonumber\\
    & \qquad {\rm IR:}\ n\cdot(+5)=5n \nonumber\\
    &{\mathbb Z}_{3n}\times [SO(n)_{\rm diag}]^2 
    \label{eq:Z3nanom}\\ 
    & \qquad {\rm UV:}\ 10n\cdot (+1)=10n \nonumber \\
    & \qquad {\rm IR:}\ n\cdot(+1)=n \nonumber
\end{align}
For the last case \eqref{eq:Z3nanom} of the discrete anomaly, the anomalies match modulo $3n$, as expected. 

Due to the symmetry breaking $SU(n)_A \times SU(n)_{\bar{F}} \rightarrow SU(n)_{\rm diag}\rightarrow SO(n)_{\rm diag}$, there are NGBs, which can be identified with the interpolating fields
\begin{align}
    &SU(n)_A \times SU(n)_{\bar{F}} / SU(n)_{\rm diag} : \nonumber\\
   & \qquad\qquad\epsilon_{abcde} \delta^{ij} \langle (A_i^{ab} A_j^{cd}) (A_k^{ef} \bar{F}^l_f) \rangle,\\
   & SU(n)_{\rm diag} / SO(n)_{\rm diag} :\nonumber\\
     & \qquad\qquad \epsilon_{abcde} \delta^k_l\langle (A_i^{ab} A_j^{cd}) (A_k^{ef} \bar{F}^l_f) \rangle.
\end{align}
For both of them, it is assumed that the trace component $\delta_k^l$ or $\delta^{ij}$ is removed. 

This symmetry breaking pattern and the massless spectrum can be ``derived'' from the tumbling hypothesis. The most attractive channel within this particle content is
\begin{align}
    \langle A_k^{5f} \bar{F}^l_f \rangle
    \propto \delta_k^l \neq 0\,,
    \label{eq:condmac}
\end{align}
which breaks the $SU(5)$ gauge group to $SU(4)$, as well as the global $SU(n)_A \times SU(n)_{\bar{F}}$ symmetry to its $SU(n)_{\rm diag}$ subgroup. The fermion representation $A$ decomposes as ${\bf 10} = {\bf 6}+{\bf 4}$, while $\bar{F}$ as $\mathbf{\bar 5} = \mathbf{\bar 4} + {\bf 1}$. The singlet components $\bar{F}^{l}_5$ remain massless, which can be identified with the composite fermions \eqref{eq:AFF}. Note that there remains an unbroken $U(1)'_{\rm PQ}$ group generated by $Q_{\rm PQ} + \frac{1}{2}(1,1,1,1,-4)$. The unbroken $SU(4)$ gauge theory is vector-like with the matter content ${n(\bf 6}_0+{\bf 4}_{-5/2}+\mathbf{\bar 4}_{5/2})$ where the subscripts are charges under $U(1)'_{\rm PQ}$. The ${\bf 6}$ representation is real and its bilinear condenses
\begin{align}
    \epsilon_{abcd} \langle A_i^{ab} A_j^{cd} \rangle \propto \delta^{ij} \neq 0\,,
\end{align}
which breaks the global $SU(n)_{\rm diag}$ to its $SO(n)_{\rm diag}$ subgroup while preserving $U(1)'_{\rm PQ}$. The other possible condensate in \eqref{eq:condmac} does break the symmetries any further. 

Given how non-trivially the 't Hooft anomaly matching conditions are satisfied, we find it very plausible that this symmetry breaking pattern and the massless particle content is correct. Thus, it is unlikely that the SU(5) model can produce a composite axion as desired.

The QCD $SU(3)_c$ part of the global symmetry is embedded as a real representation, and hence is a subgroup of the unbroken $SO(n)_{\rm diag}$. Therefore, QCD color remains unbroken. If ${\bf R}={\bf 3}+\mathbf{\bar 3}$ or ${\bf 8}$, this is indeed the case. This point, however, needs to be checked for every model depending on how $SU(3)_c$ is embedded. For both cases, the massless fermion composites are in the ${\bf 3}+\mathbf{\bar 3}$ or ${\bf 8}$ representation. For the case of ${\bf R}={\bf 3}+\mathbf{\bar 3}$, the $SU(6)_A \times SU(6)_{\bar{F}} / SU(6)_{\rm diag}$ NGBs decompose as $2({\bf 8})+{\bf 6}+\mathbf{\bar 6}+\mathbf{\bar 3}+{\bf 3}+{\bf 1}$ and the singlet remains strictly massless. 
However, for the case of ${\bf R}={\bf 8}$, the $SU(8)_A \times SU(8)_{\bar{F}} / SU(8)_{\rm diag}$ NGBs decompose as ${\bf 27}+{\bf 10}+\mathbf{\overline{10}}+2({\bf 8})$ and there are no massless singlets~\cite{Cox:2023dou}. 

Thus, given the above properties, we conclude that the SU(5) model most likely does not break $U(1)_{\rm PQ}$ and instead produces massless colored fermions (with a massless singlet scalar in the ${\bf 3}+\mathbf{\bar 3}$ case) that is phenomenologically unacceptable.

\section{A supersymmetric $SU(10)\times {\mathbb Z}_{4}$ model} 
\label{sec:SU10model}

The lesson drawn from the analysis in section~\ref{sec:GIQYmodel} is that it is important to study the dynamics of chiral gauge theories to obtain further information beyond that of 't Hooft anomaly matching. This dynamical information can then be used to argue whether the $U(1)_{\rm PQ}$ symmetry is spontaneously broken while the color $SU(3)_c$ remains unbroken in a theoretically robust fashion. 

One way to study the dynamics is to consider supersymmetric gauge theories which are perturbed with a supersymmetry-breaking parameter $m$. By using anomaly-mediated supersymmetry breaking (AMSB), which is UV insensitive, the dynamics can then be solved {\it exactly}\/ when $m\ll \Lambda$ where $\Lambda$ is the dynamical scale of the gauge theory~\cite{Murayama:2021xfj}. Concurrently, a small amount of supersymmetry breaking is welcome as a solution to the big hierarchy problem, even when $m \approx 100$~TeV. Therefore, there are two-fold motivations to study supersymmetric gauge theories.

Let us consider a supersymmetric, chiral gauge theory
where the strong gauge group is $SU(10)$ and the massless matter chiral supermultiplets transform as shown in Table~\ref{tab:matterContentSU10}. The absence of a gauge anomaly is automatic since there are six antifundamentals and one antisymmetric representation.

\begin{table*}[t]
\begin{center}
\ytableausetup{boxsize=0.7em}
\ytableausetup{aligntableaux=center}
\vspace*{-4pt}\begin{tabular}{|c|c|c||c|c||c|c|} \hline
  & $SU(10)$ & ${\mathbb Z}_{16}$   & $SU(6)_{\bar{F}}$ & $U(1)_{\rm PQ}$ & $Sp(6)_{\bar{F}}$ & $SU(3)_c\times U(1)_Y$\\
\hline \hline
$A$ & \tiny\ydiagram[]{1,1}& $+9$   & ${\bf 1}$ & $-3$ & ${\bf 1}$ & ${\bf 1}(0)$  \\[2pt] \hline
$\bar{F}$ & $\overline{\tiny\ydiagram[]{1}}$& $-4$  & \tiny\ydiagram[]{1} & $+4$ & \tiny\ydiagram[]{1} &$\tiny\ydiagram[]{1}(q_{\bar{F}})+\overline{\tiny\ydiagram[]{1}}(-q_{\bar{F}})$ \\ \hline \hline
$A\bar{F}\bar{F}$ & ${\bf 1}$& $+1$   & \tiny\ydiagram[]{1,1} & $+5$ & {\tiny\ydiagram[]{1,1}} $+{\bf 1}$& ${\bf 1}(0)+\dots$\\[2pt] \hline
${\rm Pf}A$ & ${\bf 1}$& $+13$  & ${\bf 1}$ & $-15$ & ${\bf 1}$ & ${\bf 1}(0)$ \\ \hline
\end{tabular} 
\end{center}
\caption{Representations of the chiral superfields $A,\bar{F}$ under the gauge group $SU(10)\times {\mathbb Z}_{16}$ and the global symmetry $SU(6)_{\bar{F}}\times U(1)_{\rm PQ}$. The QCD $SU(3)_c$ is embedded as ${\bf 3}+\mathbf{\bar 3}$ into $SU(6)_{\bar{F}} \supset Sp(6)_{\bar{F}} \supset SU(3)_c$. The hypercharge of $\bar{F}$ can take two possible values $q_{\bar F}=+2/3,-1/3$ in order to avoid fractionally-charged stable relics (see text for details). Representations of the $SU(10)$-invariant operators $A\bar{F}\bar{F}$ and ${\rm Pf}A$ are also shown. The missing representations of $A\bar{F}\bar{F}$ under $SU(6)\to Sp(6)\to SU(3)_c (\times U(1)_Y)$, indicated by the ellipsis, are obtained from $\mathbf{15}\to \mathbf{1}+\mathbf{14}\to
\mathbf{1}(0) + \mathbf{\bar{3}} {(2q_{\bar F})}+\mathbf{3} {(-2q_{\bar F})}+\mathbf{8}\text{(0)}$.}
\label{tab:matterContentSU10}
\end{table*}

A main feature of this model is that there are no physical $\bar \theta$-parameters. The presence of massless fermions (with at least two distinct combinations of anomalies with $SU(10)$ and $SU(3)_c$) ensures the existence of two independent anomalous chiral symmetries. This in turn allows us to absorb both $\bar \theta$-parameters and render them unphysical. The strong CP problem is then solved \emph{\`a la Peccei--Quinn}. The PQ charges in \cref{tab:matterContentSU10} follow from imposing that the PQ symmetry is $SU(10)$-anomaly free and thus it is solely broken by a $U(1)_{\rm PQ}\times \big[SU(3)_c\big]$ anomaly (see \cref{App:Discrete} for details).

The specific details of the model are given in Appendix~\ref{app:SU10dynamics} and below we summarize the crucial features for phenomenology.

%


\subsection{Symmetry breaking pattern}
Applying the exact results in \cite{Csaki:2021xhi} using AMSB, one finds that the {\it bosonic}\/ condensates which form in the near-SUSY limit $m \ll \Lambda_{10}$, with  the $SU(10)$ dynamical scale $\Lambda_{10}$, are
\begin{align}
\langle A \bar{F}^i \bar{F}^j\rangle\propto J^{ij} \neq 0\qquad \text{ and } \qquad \langle \text{Pf } A\rangle\neq 0\,,
\label{eq:bosoncond}
\end{align}
where the flavor indices $i,j=1,\dots 6$ and $J=i\sigma_2\otimes \mathbb{1}_{(N-4)/2}$ is an antisymmetric $6\times 6$ matrix. These condensates generate the following chiral symmetry breaking pattern 
\begin{align}
&SU(10)_{\rm gauge} \times SU(6)_{\bar{F}} \times U(1)_{\rm PQ} \times {\mathbb Z}_{16}\nonumber\\
  &\longrightarrow Sp(4)_{\rm gauge} \times Sp(6)_{\bar{F}}\supset SU(3)_c \times U(1)_Y\,.
\end{align}
The symmetry breaking scale (see Appendix~\ref{app:SU10dynamics}) is given by 
\begin{align}
    f_a \approx \Lambda_{10} \left(\frac{\Lambda_{10}}{m}\right)^{3/20}\,.
    \label{eq:fa}
\end{align}
At this scale the $SU(10)$ gauge symmetry is spontaneously broken to $Sp(4)$, while the global symmetry, in particular including $U(1)_{\rm PQ}$, is spontaneously broken to $Sp(6)_{\bar{F}}$ 
which contains the QCD group $SU(3)_c$. 
Below the scale $f_a$, the $Sp(4)$ gauge symmetry confines in a supersymmetric vacuum due to gaugino condensation, but the presence of AMSB at the scale $m\ll f_a$ eventually breaks supersymmetry and gives mass to the superpartners.

\subsection{Particle Spectrum}

The QCD color group $SU(3)_c$ is assumed to be embedded into the $Sp(6)_{\bar F}$ unbroken global symmetry\footnote{Note that the alternative option of embedding $SU(3)_c$ into $SU(6)$ suffers from a VEV misalignment. The VEV of an elementary Higgs transforming as $\Phi=({\bf 4}, {\bf 1}, {\bf 2})$ does not necessarily align with the VEV from the $A\bar{F}\bar{F}$ breaking.}.
Consequently, $35-21=14$ NGBs are generated for $SU(6)_{\bar{F}}/Sp(6)_{\bar{F}}$ in the anti-symmetric tensor representation of $Sp(6)_{\bar{F}}$. In the limit $\alpha_s \rightarrow 0$, the NGBs are massless, while their fermionic and (real) scalar partners obtain supersymmetry-breaking masses of ${\cal O}(m)$. However, even the NGBs acquire mass since $\alpha_s \neq 0$ explicitly breaks the global symmetry. Note that the NGBs decompose as $\mathbf{8+3+\bar 3}$ under $SU(3)_c$. Since they are all colored, the NGBs acquire masses via gluon loops of order $\frac{g_s}{2\pi} m$ where $g_s$ is the QCD gauge coupling. In particular, their masses are not raised to the dynamical scale $\Lambda_{10}$ because supersymmetry requires the pNGBs and their fermion and scalar chiral superfield partners to be split at most by ${\cal O}(m)$. Consequently, there are no massless, colored composite fermions or NGBs. Furthermore, the supersymmetry breaking scale is assumed to be $m \gtrsim 100$~TeV to be consistent with collider searches at the LHC.

Another pNGB arises through the breaking of $U(1)_{\rm PQ}$, the \emph{guaranteed composite axion.}\,\footnote{
The NGBs correspond to fluctuations around the gauge-invariant condensates in \eqref{eq:bosoncond}. Given that $f_a\gg \Lambda_{10}$, in the limit $m\rightarrow 0$, the SU(10) theory is weakly coupled at $f_a$ and the symmetry breaking scale \eqref{eq:fa} has been computed in the Higgs phase. By complementarity, this phase is continuously connected to a confined phase where the axion would be considered ``composite".} Note that it is accompanied by its axino and saxion partners with mass of ${\cal O}(m)$. 

To summarize, most of the states are at the scale $f_a$, while the following states are below or at ${\cal O}(m)$:
\begin{enumerate}
    \item The flat direction $v$ and its scalar and fermionic partners of mass ${\cal O}(m)$,
    \item The $SU(6)_{\bar{F}}/Sp(6)_{\bar{F}}$ NGBs of mass ${\cal O}(\frac{g_s}{2\pi}m)$ together with the scalar and fermionic partners of mass ${\cal O}(m)$, in
    the $\mathbf{8+3+\bar 3}$ 
    representations under $SU(3)_c$, and
    \item The almost massless axion together with its real scalar and fermionic partner of mass ${\cal O}(m)$.
\end{enumerate}
All of their mass spectra can be worked out {\it exactly}\/ despite the strong dynamics as long as $m \ll \Lambda_{10}$, even though it is unnecessary in this paper. Importantly, the $U(1)_{\rm PQ}$ symmetry is anomalous with respect to QCD and therefore the  axion will obtain the usual QCD axion mass contribution.

Note that the AMSB responsible for generating the symmetry breaking scale \eqref{eq:fa} can also be used to break supersymmetry in the minimal supersymmetric standard model (MSSM). However, as is well-known, pure AMSB would lead to tachyonic sleptons in the MSSM. To avoid this phenomenological issue, we would need to augment the supersymmetry-breaking sector with a $U(1)_R$~\cite{Pomarol:1999ie,Luty:1999qc} or $E_6$~\cite{Hook:2015tra} D-term. The details of the resulting MSSM superpartner spectrum, while interesting, will not be explored in this work.

Furthermore, above the scale $m$, the colored NGB matter contributes an amount $\Delta b_{3,\rm NGB}=4$ to the QCD gauge coupling running and $\Delta b_{1,\rm NGB}=72 q_{\bar F}^2/5$ to the hypercharge gauge coupling running.
Assuming that the SM superpartners are also at the scale $m$, this extra NGB matter contribution will upset the usual MSSM gauge coupling unification. While the QCD gauge coupling is no longer asymptotically free, there is no Landau pole below the Planck scale for $m\gtrsim 10$ TeV. To preserve unification with complete SU(5) multiplets,
the missing states could be added at the scale $m$ and may actually be part of addressing the tachyonic slepton problem. While interesting to study, we do not present further details in this work.

There is a freedom to assign electromagnetic charges to the exotic supermultiplets. If they are kept neutral the corresponding pNGBs can be dangerous since they are also coloured and would hadronize forming fractionally charged bound states which are completely stable \cite{Dunsky:2018mqs,Dunsky:2019api}. In order to avoid this issue we will charge $A$ and $\bar F$ with the same hypercharge of either the down or up quarks, see \cref{tab:matterContentSU10}. This will fix the anomaly coefficients for the $U(1)_{\rm PQ}$ that determine the axion couplings.
\begin{align}
 \mathcal{L}_{a}&\supset
 \frac{\alpha_{EM}}{8\pi}\frac{E}{N}\frac{a}{f_{a}} F_{\mu\nu} \tilde F^{\mu\nu} + \frac{\alpha_s}{8\pi} \frac{a}{f_a}G_{a\,\mu\nu} \tilde G_a^{\mu\nu}\,.
 \label{Eq:AxionEff}
 \end{align}
 Thus the axion-photon coupling will have a fixed $E/N$ ratio among these two possible values, namely
\begin{align}
E/N=\Bigg\{
\begin{array}{rll}
2\times 1/3\quad  &\text{for} \quad q_{\bar{F}}=-1/3\,,\\
2\times 4/3\quad  &\text{for} \quad q_{\bar{F}}=2/3\,.
\end{array}
\end{align}
where we have defined $f_a\equiv f_{\rm PQ}/N$ and the anomaly factors read
$E=2 \sum_f q_{{\rm PQ},f} \, q_{{\rm EM},f}^2; N= 2\sum_f q_{{\rm PQ},f}\,  T(R_{{\rm QCD},f})$, where $q_{{\rm PQ},f}$ $(q_{{\rm EM},f})$ is the PQ (electric) charge of the fermion $f$,   
and $T(R_{{ \rm QCD},f})$ is the Dynkin index of the corresponding $SU(3)_c$ representation, defined as $T(R) \delta_{a b}\equiv \operatorname{Tr}\left(t_R^a t_R^b\right)$, with 
$T(\tiny\ydiagram{1})=1/2$.

This charge assignment can also be understood by taking into account the global structure of the SM gauge group. The minimal SM gauge group is $SU(3)\times SU(2)\times U(1)/{\mathbb{Z}}_6$, where ${\mathbb{Z}}_6$ is generated by the product of $e^{2\pi i/3}$ on color triplets, $-1$ on weak doublets, and $e^{2\pi iY}$ on hypercharges (see, e.g., \cite{Tong:2017oea}). The SM fields are invariant under this ${\mathbb{Z}}_6$, and as long as the exotic matter content remains invariant, no fractionally charged states will arise in the IR.

\subsection{Axion quality problem} 

When the gauge theory is coupled to gravity, the Peccei-Quinn global symmetry can be violated and potentially misalign the axion potential. The possible PQ-violating terms are dictated by the gauge and Lorentz symmetry. In particular, the superpotential can include a term breaking $U(1)_{\rm PQ}$ (by $+5$) at the renormalizable level,
\begin{align}
W_{\text{tree}}= 
    J_{ij} A \bar{F}^i \bar{F}^j .
\end{align}
Interestingly, this term can be forbidden by ``gauging" the discrete symmetry ${\mathbb Z}_{16}$ in Table \ref{tab:matterContentSU10} or one of its subgroups, analogous to how baryon and lepton number violation 
in the MSSM can be suppressed with discrete gauge symmetries~\cite{Ibanez:1991hv,Ibanez:1991pr,Ibanez:1992ji}. In fact the discrete symmetry ${\mathbb Z}_{16}$ is not arbitrary and dictated by the anomaly-free condition under the $SU(10)$ instanton. 
The cancellation of the discrete anomalies is discussed in Appendix~\ref{App:Z16anomalies}.

Assuming a ${\mathbb Z}_{16}$ discrete symmetry, the lowest dimension operator that violates $U(1)_{\rm PQ}$ 
is the dimension 28 Kahler term
\begin{align}
K\supset \frac{c_{\tiny\cancel{PQ}}}{M_{\rm P}^{26}} ({\rm Pf}A)^{*5} (A \bar{F} \bar{F})\,.
\label{Eq_PQVZ16}
\end{align}
This compares to the lowest dimension superpotential term which has dimension 36.
If only the ${\mathbb Z}_8$ subgroup is imposed as a symmetry, the lowest dimension operator is dimension 16
\begin{align}
K\supset \frac{c_{\tiny\cancel{PQ}}}{M_{\rm P}^{14}} ({\rm Pf}A)^{*2} (A \bar{F} \bar{F})^2\,,
\label{Eq_PQVZ8}
\end{align}
while the lowest dimension superpotential term has dimension 18.
Finally, if only the ${\mathbb Z}_4$ subgroup is imposed the lowest dimension PQ-violating operator has dimension 8:
\begin{align}
K\supset \frac{c_{\tiny\cancel{PQ}}}{M_{\rm P}^{6}} ({\rm Pf}A)^{*} (A \bar{F} \bar{F})\,.
\label{Eq_PQVZ4}
\end{align}

Instead, the lowest dimension superpotential term has dimension 12.

These PQ-breaking operators induce a displacement of the axion vacuum expectation value from the CP conserving minimum. The non-zero induced displacement must satisfy $\Delta\bar \theta_{\rm eff}\lesssim 10^{-10}$ to comply with the experimental bounds on the neutron electric dipole moment \cite{Pendlebury:2015lrz,Baker:2006ts}. For each of the PQ breaking operators in \cref{Eq_PQVZ16,Eq_PQVZ8,Eq_PQVZ4} suppressed by 
$1/M_P^n$, the corresponding displacement induces a bound on $f_a$,
\begin{align}
    \Delta\bar \theta_{\rm eff}=
   c_{\tiny\cancel{PQ}} \frac{m^2f_a^{n+2}}{\chi_{\rm QCD} \, M_P^n} \frac{2^{19}\cdot 5^8}{3^3\cdot 7^4} < 10^{-10}\,,
    \label{Eq:displacement}
\end{align}
where $\chi_{\rm QCD}\simeq m_\pi^2 f_\pi^2 \frac{m_u m_d}{(m_u+m_d)^2}$  is the QCD topological susceptibility. The maximal values of $f_a$ consistent with the bound in \cref{Eq:displacement} for each discrete symmetry are
\begin{align}
\mathbb{Z}_4:\quad &f_a < 3.7\times 10^{10}\,\text{GeV},\nn\\
\mathbb{Z}_8:\quad &f_a < 6.7\times 10^{14}\,\text{GeV},\nn\\
\mathbb{Z}_{16}:\quad &f_a < 4.5\times 10^{16} \,\text{GeV},
\end{align}
assuming the overall coefficient $c_{\tiny\cancel{PQ}}\sim 0.1$
These bounds are also shown in \cref{fig:photons}.

As discussed in Appendix~\ref{App:Z16anomalies}, the discrete gravitational anomalies are not automatically cancelled. However, this does not prevent gauging the discrete symmetry because spectator fields can be introduced to cancel these anomalies. For instance, to cancel the ${\mathbb Z}_{16}\times [grav]^2$ anomaly we can introduce three extra $SU(10)$ gauge singlets with ${\mathbb Z}_{16}$ charge $+1$. Of course, these singlet fields allow much lower dimension 
PQ-violating operators, but provided these fields do not receive a VEV they will not lead to a misalignment of the QCD axion potential.

\begin{figure}[t]
\centering
\includegraphics[scale=0.25]{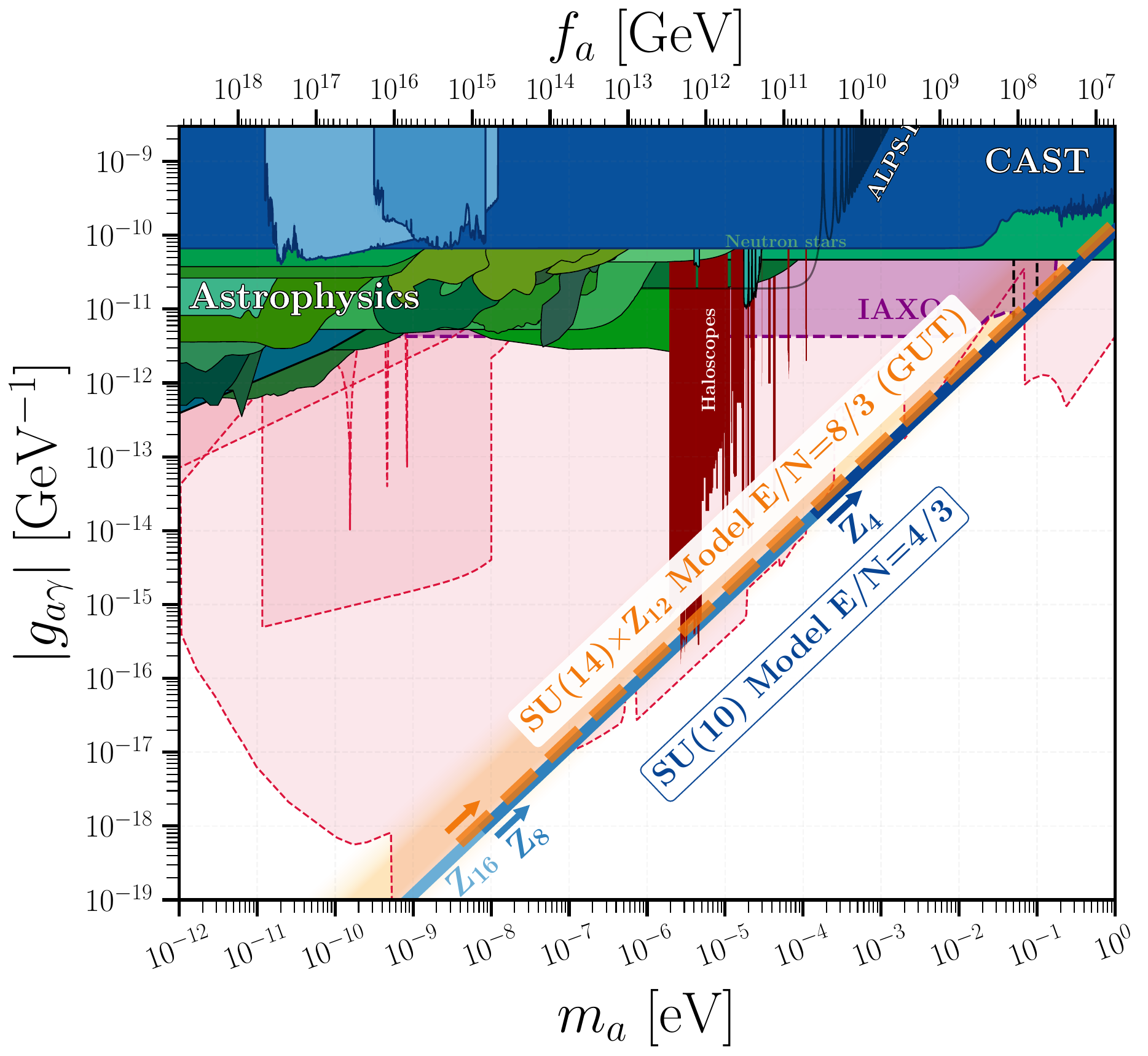}
\caption{Axion limits adapted from Ref.~\cite{AxionLimits}. 
}
\label{fig:photons}
\end{figure}
\subsection{Collider signatures}

The lightest particles beyond the axion itself are the colored NGBs which leads to several interesting collider signatures. The NGBs acquire mass from QCD at the one-loop level $\sim\frac{g_s}{2\pi} C_i m$ and thus are lighter than the supersymmetry breaking scale $m$. The immediate prediction is that the triplet and octet have the mass ratio of $4:9$ by their quadratic Casimirs $C_i$. 
The color-octet NGBs decay to $gg$ via the loop of heavy vector multiplets. Their lifetimes can vary from prompt for $f_a \approx 10^9$~GeV to displaced for $f_a \approx 10^{12}$~GeV. Limits on promptly decaying color-octet NGBs decaying to gluons are excluded below 800 GeV using the scalar gluon search in Ref.~\cite{ATLAS:2017jnp}. It is within the range of masses that can be explored at 10~TeV pCM colliders in the future. 

On the other hand, assuming the ${\mathbb Z}_{4}$ discrete symmetry, the triplet NGB can decay only via the Planck-suppressed operator such as
\begin{align}
   K &= \frac{1}{M_{\rm P}^{8}} ({\rm Pf}A)^* (A\bar{F}^i\bar{F}^j) d^c_i d^c_j \,,
\end{align}
when $\bar{F}^i$ has color ${\bf 3}$ and hypercharge $-1/3$ (replace $d^c$ by $u^c$ if the hypercharge is $+2/3$). Due to the extremely large $M_{\rm P}$ suppression, the lifetime is exceedingly longer than the age of the universe. When it is produced at a future collider, it becomes a stable heavy hadron. There are two possibilities depending on the hypercharge assignment. If $q_{\bar F} =-\frac{1}{3}$ then $A\bar{F}\bar{F}$ contains the charged states ${\bar X}({\bar{\bf 3}},-\frac{2}{3})+X({\bf 3},\frac{2}{3})$ and $X$ would bind into $(X\bar{u})^0$ which is electromagnetically neutral. In this case, heavy hydrogen would not form, avoiding bounds from heavy hydrogen searches in sea water. Similarly, if $q_{\bar F} =\frac{2}{3}$ then $A\bar{F}\bar{F}$ contains the charged states ${\bar X}({\bar{\bf 3}},\frac{4}{3})+X({\bf 3},-\frac{4}{3})$. These states could form the neutral hadron $(X u u)^0$ or the electromagnetically charged hadron $(X u d)^-$. The charged hadrons lead to an anomalously large ionization energy loss ($dE/dx$), and hits in the hadron calorimeter at the LHC while the neutral hadrons only have hits in the calorimeters. Current limits on long-lived sbottoms from R-hadron searches at the LHC, can be used to constrain the color triplet $X$ to be $\gtrsim 1.25$ TeV~\cite{ATLAS:2019gqq}. The color-octet pNGB would decay into $gg$ in a way analogous to $\pi^0 \rightarrow \gamma\gamma$ in QCD due to the chiral anomaly. It may have a displaced vertex signature depending on $f_a$ and $m$.


\subsection{Relics}

Colored pNGBs are produced in the early universe if the reheating temperature after inflation is above their masses. If a hadron made of a stable charged triplet pNGB would combine with an electron (either by itself or together with a proton) to form a superheavy hydrogen isotope, it can be looked for in experiments. Searches for such a heavy hydrogen-like isotope in seawater require halo fractions below $10^{-2}-5\times 10^{-5}$ for positively charged 100 TeV hadron masses~\cite{Verkerk:1991jf}, which is the abundance expected for a $\sim 10$~GeV colored pNGB from the thermal freeze-out with perturbative QCD.
On the other hand, when the temperature comes down to the QCD scale, interactions become strong and $X$ may recouple and annihilate further and get depleted \cite{DeLuca:2018mzn,Cirelli:2024ssz}. It would be interesting to study this question further.

If the perturbative thermal freeze-out is the main driver for determining the abundance, it may be subject to various observational and experimental constraints. If we use the option of $q_F=-1/3$ for color triplets in $\bar{F}_{\bf 3}$, the color-triplet pNGB $X^i=\epsilon_{ijk}\bar{F}_{\bf 3}^j\bar{F}_{\bf 3}^k$ has $Y=Q=+2/3$. Then its bound state $X_u^0=X\bar{u}$ is electrically neutral and would not form a heavy hydrogen isotope. The charged state $X_d^+=X\bar{d}$ would $\beta$-decay to this neutral state $X_d^+ \rightarrow X_u^0 e^+ \nu_e$ before BBN because it is heavier, similar to $m_{D^+} - m_{D^0} =  4.82$ MeV. Unless $X_u$ would be bound to nuclei, it is not subject to this constraint. A baryon-like state $\Lambda_X^{+} = X u d$ would be a concern but presumably it is a small fraction of $X$. If we use the option of $q_F=+2/3$ instead, the baryon-like state $\Lambda_X^0$ is neutral, while the meson-like state $X_u^-$ is not. Given the higher probability of forming mesons than baryons, it is subject to a stronger constraint. In either case, the abundance of heavy isotopes may be in the interesting range. In addition, there are interesting and relevant constraints from balloon-borne experiments with plastic tracking detectors \cite{Starkman:1990nj}. There do not appear to be more recent constraints of this type. It would be another interesting direction for experimental searches for dark matter.

\section{A supersymmetric $SU(14)\times {\mathbb Z}_{12}$ model} 
\label{sec:SU10model}

The $SU(10)$ model in section~\ref{sec:SU10model} can be extended to an $SU(14)$ gauge theory in order to incorporate the grand unified groups such as $SU(5)$ and $SO(10)$, which can weakly gauge the global symmetry. The chiral superfield matter content consists of ten antifundamentals ${\bar F}_i ~(i=1,\dots,14)$ and one antisymmetric representation $A$ of $SU(14)$ where the $U(1)_{\rm PQ}$ symmetry is again associated with one of the independent $U(1)$ symmetries that is $SU(14)$ anomaly-free. The representations and PQ charges are shown in Table~\ref{tab:matterContentSU10}.

The nonzero condensates $\langle A \bar{F}^i \bar{F}^j\rangle$ and $\langle \text{Pf } A\rangle$ spontaneously break $SU(14)_{\rm gauge}\rightarrow Sp(4)_{\rm gauge}$ and the global symmetry $SU(10)_{\bar F}\rightarrow Sp(10)_{\bar F}$, at the scale $f_a$ (see \eqref{eq:faSU14}). 
The axion is again identified with the spontaneous breaking of the $U(1)_{\rm PQ}$.
There are two options for how the grand unified groups are embedded into the global symmetry $SU(10)_{\bar F}$. In the first option, we can weakly gauge an $SU(5)$ subset of the unbroken global symmetry $Sp(10)_{\bar F}$. The resulting global symmetry breaking gives $99-55=44$ NGBs which decompose as $\mathbf{24+10+\bar{10}}$ under the weakly-gauged $SU(5)$. 
The charged NGBs obtain a mass of ${\cal O}(\frac{g_{\rm GUT}}{2\pi}m)$ where $g_{\rm GUT}$ is the $SU(5)$ gauge coupling at the scale $f_a$, while the scalar and fermionic partners obtain a supersymmetry breaking mass of ${\cal O}(m)$. The extra $SU(5)$ matter fields lead to Landau poles below the GUT scale unless $m\gtrsim 10^{12}$ GeV.

A second option is to weakly gauge $SO(10) \subset SU(10)_{\bar F}$. The global symmetry breaking now spontaneously breaks the weakly-gauged $SO(10)$ to $SU(5)\times U(1)$ since $SO(10) \cap Sp(10)_{\bar F} \simeq U(5)$ is the unbroken symmetry. This means $45-25= 20$ NGBs are eaten which are precisely the $\mathbf{10+\bar{10}}$. Thus, the only remaining NGBs transform as $\mathbf{24}$ of $SU(5)$. Again these charged NGBs obtain a mass of ${\cal O}(\frac{g_{\rm GUT}}{2\pi}m)$, while the scalar and fermionic partners obtain a supersymmetry breaking mass of ${\cal O}(m)$. With this reduced NGB matter content there are no longer any Landau poles if the supersymmetry breaking scale $m=100$ TeV and the decay constant (or unification scale), $f_a\simeq 10^{16}$ GeV. 
An additional elementary ${\bf 24}_H$ and ${\bf 126}_H$ (or ${\bf 16}_H$) is then assumed to break $SU(5)\times U(1)$ to the SM gauge group.

To avoid PQ-violating terms at the renormalizable level, we proceed as in the SU(10) example in section~\ref{sec:SU10model}.
The largest discrete group $\Gamma$ of the $SU(14)$ model that is free from  $\Gamma\times \left[SU(10)\right]^2$ and $\Gamma\times \left[SU(5)\right]^2$ (or $\Gamma\times \left[SO(10)\right]^2$) mixed anomalies is $\Gamma=\mathbb{Z}_{24}$, which can therefore be consistently gauged. To obtain a high quality axion, it suffices to only gauge a $\mathbb{Z}_{12}$  subgroup of $\Gamma$, with charge assignments given in Table~\ref{tab:matterContentSU14}.
With this ${\mathbb Z}_{12}$ discrete symmetry, the lowest dimension operator that violates $U(1)_{\rm PQ}$ 
is the dimension 20 Kahler term
\begin{align}
K\supset \frac{c_{\tiny\cancel{PQ}}}{M_{\rm P}^{18}} ({\rm Pf}A)^{*2} (A \bar{F} \bar{F})^2\,.
\label{Eq_PQVZ12}
\end{align}
This compares to the lowest dimension superpotential term which has dimension 30. The operator \eqref{Eq_PQVZ12} allows the bound on the axion decay constant scale $f_a\lesssim 1.5\times 10^{15}$ GeV.  This is marginally compatible with MSSM unification which occurs at $\sim 2\times 10^{16}$ GeV. The discrete symmetry can be increased to ${\mathbb Z}_{24}$ where the lowest dimension operator is now a dimension 38 superpotential term which gives the upper bound $f_a\lesssim 4.5\times 10^{16}$ GeV.  This is nicely compatible with MSSM unification.
However, note that a large axion decay constant scale $f_a\gtrsim 10^{11}$ GeV, typically overproduces the axion dark matter via the misalignment mechanism, unless the initial misalignment angle is tuned, which may be anthropic. Moreover, there could also be a dark matter component from the neutralino spectrum, and the interplay between the axion and the WIMP mass would be interesting to investigate further.

Finally, in both grand unified scenarios, the ratio $E/N$ entering the axion--photon coupling is uniquely determined by the Standard Model embedding in the GUT group~\cite{Srednicki:1985xd,Agrawal:2022lsp}, yielding $E/N = 8/3$. This provides a well-defined target for axion experiments (see \cref{fig:photons}).

\begin{table*}[t]
\begin{center}
\ytableausetup{boxsize=0.7em}
\ytableausetup{aligntableaux=center}
\vspace*{-4pt}\begin{tabular}{|c|c|c||c|c||c|c|} \hline
  & $SU(14)$ & ${\mathbb Z}_{12}$   & $SU(10)_{\bar{F}}$ & $U(1)_{\rm PQ}$ & $Sp(10)_{\bar{F}}$ & $SU(3)_c\times U(1)_Y$\\
\hline \hline
$A$ & \tiny\ydiagram[]{1,1}& $+1$   & ${\bf 1}$ & $-5$ & ${\bf 1}$ & ${\bf 1}(0)$  \\[2pt] \hline
$\bar{F}$ & $\overline{\tiny\ydiagram[]{1}}$& $+6$  & \tiny\ydiagram[]{1} & $+6$ & \tiny\ydiagram[]{1} &$\overline{\tiny\ydiagram[]{1}}(+1/3)+{\bf 1}(-1)+\text{c.c.}$ \\ \hline \hline
$A\bar{F}\bar{F}$ & ${\bf 1}$& $+1$   & \tiny\ydiagram[]{1,1} & $+7$ & {\tiny\ydiagram[]{1,1}} $+{\bf 1}$& ${\bf 1}(0)+\dots$\\[2pt] \hline
${\rm Pf}A$ & ${\bf 1}$& $+7$  & ${\bf 1}$ & $-35$ & ${\bf 1}$ & ${\bf 1}(0)$ \\ \hline
\end{tabular} 
\end{center}
\caption{Representations of the chiral superfields $A,\bar{F}$ under the gauge group $SU(14)\times {\mathbb Z}_{12}$ and the global symmetry $SU(10)_{\bar{F}}\times U(1)_{\rm PQ}$. 
Representations of the $SU(14)$-invariant operators $A\bar{F}\bar{F}$ and ${\rm Pf}A$ are also shown. 
}
\label{tab:matterContentSU14}
\end{table*}
\section{Conclusion}

Using supersymmetric chiral dynamics we have constructed a model with a high-quality, composite axion that solves the strong CP problem. In a minimal model, this is done by perturbing a supersymmetric $SU(10)$ chiral gauge theory with anomaly-mediated supersymmetry breaking where the QCD gauge group weakly gauges the unbroken global $Sp(6)$ symmetry. This enables the nonsupersymmetric vacuum to be calculated exactly, where the theory spontaneously breaks an anomalous Peccei-Quinn symmetry, giving rise to a NGB which is identified with the QCD axion. Due to the supersymmetric strong dynamics the axion decay constant can be calculated exactly as a function of the $SU(10)$ strong coupling scale. Furthermore, given that the anomaly-mediated supersymmetry breaking scale can be near 100 TeV helps to address the big hierarchy problem, that is normally ignored in axion model-building.

To protect the PQ global symmetry from quantum-gravity effects we identify a discrete $\mathbb{Z}_4$ symmetry that is anomaly-free and therefore can be gauged. This discrete symmetry forbids PQ-violating operators up to dimension 8 corresponding to $f_a\lesssim 3.7\times 10^{10}$ GeV or axion-photon couplings $|g_{a\gamma}| \lesssim  10^{-14}$. This provides a reachable target for future axion experiments. Interestingly, the composite operators from the strong dynamics helps to reduce the order $N$ of the discrete $\mathbb{Z}_N$ symmetry compared to supersymmetric models with only elementary states (e.g., see \cite{Babu:1989rb}). The PQ symmetry can be protected to even higher order with $\mathbb{Z}_8$ and $\mathbb{Z}_{16}$ discrete symmetries which can also be gauged. 

Besides the axion, there are colored NGBs that could provide an experimental signal of the strong dynamics. The NGBs decay to gluons either promptly or displaced that can be detected at future colliders. If the NGBs are very long-lived they may even form stable hadrons which leave distinctive collider signals. The stable hadrons containing a charged triplet pNGB could also form a superheavy hydrogen isotope in the early universe that may have the appropriate mass to be detected in balloon-borne or terrestrial experiments.

The minimal $SU(10)$ model can be extended to an $SU(14)$ model that incorporates grand unification via an $SU(5)$ or $SO(10)$ group. In particular, the $SO(10)$ unified group weakly gauges the $SU(10)$ global symmetry. The strong dynamics dynamically breaks $SO(10)$ to $SU(5)\times U(1)$ giving rise to NGBs transforming as $\bf 24$, besides the QCD axion. Given that the supersymmetry breaking scale is near 100 TeV, gauge coupling unification is preserved, and thus the unification scale $\sim 10^{16}$ GeV can be identified with the axion decay constant scale. This value for the axion decay constant requires a tuning in the initial misalignment angle to prevent axion dark matter from overclosing the universe. The neutralino could also contribute to the dark matter abundance and exploring the possible parameter space would be interesting to further study. Moreover, a discrete $\mathbb{Z}_{12}$ gauge symmetry can be identified which forbids PQ-violating operators up to dimension 20.

Supersymmetric chiral gauge theories perturbed by anomaly-mediated supersymmetry breaking provide a novel framework in which to address the strong CP and axion quality problem as well as the Higgs hierarchy problem and possible connections to grand unification. It would be interesting to further explore whether these theories with exact solutions admit other possibilities for model-building that address the puzzles of the Standard Model.

\vspace{1cm}
\noindent
{\bf Note added:} While this paper was in preparation we became aware of an independent study~\cite{Sato}
that also obtains a QCD axion from a supersymmetric chiral gauge theory.

\section*{Acknowledgements}
We thank David Dunsky and Xiaochuan Lu for useful discussions.
The work of T.\,G. is supported in part by the Department of Energy under Grant No. DE-SC0011842 at the University of Minnesota. The work of H.\,M.\ is supported by the Director, Office of Science, Office of High Energy Physics of the U.S. Department of Energy under the Contract No. DE-AC02-05CH11231, by the NSF grant PHY-2515115, by the JSPS Grant-in-Aid for Scientific Research JP20K03942, Hamamatsu Photonics, K.K, and Tokyo Dome Corporation. In addition, H.\,M.\ is supported by the World Premier International Research Center Initiative (WPI) MEXT, Japan.
The work of P.\,Q. is supported by the U.S. Department of Energy under grant number DE-SC0009919 and partially supported by the European Union's Horizon 2020 research and innovation programme under the Marie Sk\l odowska-Curie grant agreement No 101086085-ASYMMETRY.
Likewise, H.\,M., T.\,G. and P.\,Q. thank the CERN theory group for their warm hospitality during the \emph{Crossroads between Theory and Phenomenology} Program, where this work was initiated.
\appendix
\section{Details of the dynamics}
\subsection{The $SU(10)$ case}
\label{app:SU10dynamics}

In the supersymmetric limit, the $SU(10)$ theory introduced in section~\ref{sec:SU10model} is known to have ``run-away" behavior, namely that the non-perturbative dynamics forces the bosonic fields to turn on and run to the infinite expectation values. The derivation starts from the observation that the scalar (or $D$-term) potential has the form
\begin{align}
    V_D &= \frac{1}{2} g_{10}^2 \sum_a
    ({\rm Tr}\, T^a (A A^\dagger - \bar{F}^* \bar{F}^T))^2\,,
\end{align}
where the sum is over the $SU(10)$ generators.
This potential admits a flat direction in field space given by (zero entries are not written to avoid clutter)
\begin{align}
    A &= \left( \begin{array}{cccccc|cccc}
     & a &  &  &  &  &  &  &  &  \\
    -a &  &  &  &  &  &  &  &  &  \\
     &  &  & a &  &  &  &  &  &  \\
     &  & -a &  &  &  &  &  &  &  \\
     &  &  &  &  & a &  &  &  &  \\
     &  &  &  & -a &  &  &  &  &  \\ \hline
     &  &  &  &  &  &  & b &  &  \\
     &  &  &  &  &  & -b &  &  &  \\
     &  &  &  &  &  &  &  &  & b \\
     &  &  &  &  &  &  &  & -b & 
    \end{array} \right),\\  
    \bar{F} &= \left( \begin{array}{cccccccccc}
    c &  &  &  &  &   \\
     & c &  &  &  &   \\
     &  & c &  &  &   \\
     &  &  & c &  &   \\
     &  &  &  & c &  \\
     &  &  &  &  & c  \\ \hline
     &  &  &  &  &   \\
     &  &  &  &  &   \\
     &  &  &  &  &   \\
     &  &  &  &  &  
    \end{array} \right)\,,
\end{align}
where the $D$-flat condition ($V_D=0$) implies that $|a|^2 = |b|^2 + |c|^2$. Along this direction, the gauge group is broken from $SU(10)$ to $Sp(4)$. The low-energy theory is a pure supersymmetric Yang--Mills theory ({\it i.e.}\/, no matter fields) which is known to confine and produce a gaugino condensate with the superpotential $\Lambda_{Sp(4)}^3$. By simply matching the low-energy and high-energy gauge couplings, we obtain the non-perturbative superpotential\footnote{Note that we are absorbing any overall coefficient into the definition of the dynamical scale $\Lambda_{10}$.}
\begin{align}
    W &= \Lambda_{Sp(4)}^3
    = \left(
    \frac{\Lambda_{10}^{23}}{({\rm Pf}A) ({\rm Pf} A \bar{F}\bar{F})} \right)^{1/3}\,,
\end{align}
where $\Lambda_{10}$ is the $SU(10)$ dynamical scale.
In fact, this is the only superpotential that is consistent with all the symmetries of the theory and hence is {\it exact}\/. It produces a potential $V_{\rm SUSY} \sim \left|\Lambda_{10}^{23}/v^{17} \right|^{2/3}$ (where $v$ is the characteristic scale of the moduli) that indeed has a minimum that runs away to infinity i.e. $v \rightarrow \infty$. 

However, in the presence of a small supersymmetry breaking $m$, due to anomaly mediation, there is an additional term in the potential given by~\cite{Murayama:2021xfj}
\begin{align}
    V_{\rm AMSB} = m \left(\varphi_i \frac{\partial W}{\partial\varphi_i}-3W\right) + c.c.\,,
    \label{eq:VAMSB}
\end{align}
where $\varphi_i$ denote generic scalar fields.
Adding the contribution \eqref{eq:VAMSB} to $V_{\rm SUSY}$ leads to the potential
\begin{align}
    V =& \frac{4}{3} \left| \frac{\Lambda_{10} ^{23/3}}{a^3 b^{2/3} c^2}\right| ^2+\frac{2}{3} \left| \frac{\Lambda_{10} ^{23/3}}{a^2 b^{2/3} c^3}\right| ^2\nonumber\\
    &+\frac{2}{9} \left| \frac{\Lambda_{10} ^{23/3}}{a^2 b^{5/3} c^2}\right| ^2-\frac{23 \Lambda_{10} ^{23/3} m}{3 a^2 b^{2/3} c^2} + c.c.
\end{align}
The theory now has a stable ground state at
\begin{align}
    b=c=a/\sqrt{2} &= 
    \Lambda_{10}\left(\frac{17\,\Lambda_{10}}{138\,m}\right)^{3/20} ,
\end{align}
and provided that $m \ll \Lambda_{10}$, the solution is {\it exact}\/. 

The value of the moduli at the minimum can be related to the PQ-breaking scale $f_{PQ}$ by expanding the scalar kinetic terms.  By substituting
\begin{align}
    A &= \left( \begin{array}{c|c} a J_6 & 0 \\ \hline 0 & b J_6 \end{array} \right) e^{-3ia/f_{PQ}}, \\
    \bar{F} &= \left( \begin{array}{c} c I_6 \\ \hline  0 \end{array} \right) e^{+4ia/f_{PQ}},
\end{align}
the kinetic terms become 
\begin{align}
    {\cal L}_K &= \frac{1}{2} {\rm Tr} \partial A^\dagger \partial A
    + {\rm Tr} \partial \bar{F}^\dagger \partial \bar{F} \nonumber \\
    & = \frac{1}{f_{PQ}^2} ((-3)^2(3|a|^2+2|b|^2)+4^2 6|c|^2) \partial a \partial a \nonumber \\
    & = \frac{1}{2}\partial a \partial a,
\end{align}
such that the axion is canonically normalized for
\begin{align}
    f_{PQ}^2 &= 2(27|a|^2 + 18|b|^2 + 96|c|^2).
\end{align}
Then the axion decay constant $f_a = f_{PQ}/N$, where the color anomaly factor is $N = 2 \sum q_{PQ} T(R_{QCD}) = 80$ for this theory. We then obtain
\begin{align}
    f_a =& \frac{\sqrt{42}}{40}a \approx 0.17\, \Lambda_{10}\left(\frac{\Lambda_{10}}{m}\right)^{3/20}\,.
\end{align}

\subsection{The $SU(14)$ case}
For the $SU(14)$ theory, the analogous calculation gives the minimum
\begin{align}
    b=c=a/\sqrt{2}=& \Lambda_{14}\left(\frac{25\,\Lambda_{14}}{2^{5/3} 93\,m}\right)^{3/28}\,,
\end{align}
and thus, using the PQ charges $q_A = -5$ and $q_{\bar{F}}=+6$ from \cref{tab:matterContentSU14} we obtain
\begin{align}
    f_{PQ}^2 =&  2\left(125\abs{a}^2+ 50\abs{b}^2+360\abs{c}^2\right)\,,
    \\
    f_{PQ}\approx&\, 27.9\, \Lambda_{14}\left(\frac{\Lambda_{14}}{m}\right)^{3/28}\,.
\end{align}
The axion decay constant is then calculated to be
\begin{align} 
    f_a =& \frac{f_{PQ}}{N} \approx  0.17\, \Lambda_{14}\left(\frac{\Lambda_{14}}{m}\right)^{3/28}\,,
    \label{eq:faSU14}
\end{align}
where the color anomaly coefficient is $N= 168$ in this theory.

\section{Anomaly-free discrete symmetry of the $SU(10)$ model}
\label{App:Discrete}

In this appendix we derive  the largest discrete symmetry $\Gamma$ of the $SU(10)$ model, which is exact at the renormalizable level and respected by both the $\Gamma\times \left[SU(3)_c\right]^2$ and $\Gamma\times \left[SU(10)\right]^2$ mixed anomalies. This discrete symmetry needs to be contained in an arbitrary $U(1)_{A}\times U(1)_{\bar{F}}$ rotation of the $A,\bar{F}$ superfields, which can be arranged in such a way that one $U(1)$ (defined as $U(1)_{\rm PQ}$) is only anomalous under $SU(3)_c$, while the other $U(1)$ (denoted by $U(1)'$) is only anomalous under $SU(10)$. This leads to the anomaly conditions
\begin{align}
U(1)_{\rm PQ}\times \left[SU(10)\right]^2:& \quad 0=6\, q_{{\rm PQ},\bar{F}}+8 \, q_{{\rm PQ},A}\nn \\
U(1)'\times \left[SU(3)_c\right]^2:& \quad 0=10\, q'_{\bar{F}}\,. 
\label{Eq:anoamliesGenG}
\end{align}
\begin{table}[h]
\begin{align*}
\begin{array}{|c|c|c||c|c|} 
\hline
                   & U(1)_{\rm PQ} & {\mathbb Z}_{80}   &  U(1)' & {\mathbb Z}'_8 \\
\hline A           &  -3 &  -3 & +1  & +1 \\
\hline \bar{F}     &  +4  & +4 &  0  & 0 \\
\hline\hline       
A \bar{F} \bar{F}  &  +5 &   +5  &  +1  & +1 \\
\hline {\rm Pf } A &  -15 &   -15 &  +5 &  +5 \\
\hline   
\end{array}
\end{align*}
\caption{Charge assignments of the anomalous abelian symmetries $U(1)_{\rm PQ}\times U(1)'$ and their corresponding anomaly-free discrete subgroup ${\mathbb Z}_{80}\times {\mathbb Z}_{8}'$.
\label{Tab:Discrete charges}}
\end{table}

The first equation in \eqref{Eq:anoamliesGenG} determines 
the corresponding PQ charges of $A$ and $\bar{F}$, while imposing  
the second equation in \eqref{Eq:anoamliesGenG} yields the corresponding $U(1)'$ charges, as shown in \cref{Tab:Discrete charges}. Both abelian symmetries are anomalous under the other gauge group. In particular, $U(1)_{\rm PQ}$ exhibits a QCD anomaly, while $U(1)'$ has a $SU(10)$ anomaly, given by
\begin{align}
U(1)_{\rm PQ}\times \left[SU(3)_c\right]^2:& \quad (+4) \cdot 2 \cdot 10 \cdot 2T(\bar{F})=80\,, \nn\\ 
U(1)'\times \left[SU(10)\right]^2:& \quad (+1)\cdot 2T(\bar{A})=8\,.
\label{Eq:anoamliesPQandprime}
\end{align}
However, a discrete subgroup of each $U(1)$ symmetry will remain anomaly free, since the anomaly coefficients in \cref{Eq:anoamliesPQandprime} are larger than one.
The candidates for the discrete symmetry $\Gamma$ are a ${\mathbb Z}_{80}$ subgroup of $U(1)_{\rm PQ}$ and a ${\mathbb Z}_{8}'$ subgroup of $U(1)'$, whose charges are shown in \cref{Tab:Discrete charges}.
Note that it is important to consider that some elements of ${\mathbb Z}_{80}\times {\mathbb Z}_{8}'$, belong to the center ${\mathbb Z}_{10}$ of the $SU(10)$ group.   Consequently, these redundant transformations must be factored out to obtain the discrete group that imposes non-trivial constraints on gauge-invariant operators. First of all, one can show that any transformation in ${\mathbb Z}_{80}$ can be obtained by a combined ${\mathbb Z}_{16}$ 
and ${\mathbb Z}_{10}$ transformation, as follows
{\fontsize{10pt}{11pt}\selectfont
\begin{align}
   &A\xrightarrow{ {\mathbb Z}_{80}}e^{-3\frac{2\pi i}{80}}A;\,\,\nn\\& A\xrightarrow{ {\mathbb Z}_{16}\times  {\mathbb Z}_{10}}e^{(9\frac{2\pi i}{16}+2\cdot2\frac{2\pi i}{10} )}A =e^{-3\frac{2\pi i}{80}}A\,.\nn\\
   & \bar{F}\xrightarrow{ {\mathbb Z}_{80}}e^{+4\frac{2\pi i}{80}}\bar{F};\,\,\nn\\& \bar{F}\xrightarrow{ {\mathbb Z}_{16}\times  {\mathbb Z}_{10}}e^{(4\frac{2\pi i}{16} -1\cdot 2\frac{2\pi i}{10})}\bar{F} =e^{+4\frac{2\pi i}{80}}\bar{F}\,.
   \label{Eq:Z80toZ16} 
\end{align}}
where the charges are given in \cref{Tab:Discrete charges}.
Similarly, any element in ${\mathbb Z}_{8}'$, can also be obtained by a combination of the ${\mathbb Z}_{16}$ in \eqref{Eq:Z80toZ16} and the center ${\mathbb Z}_{10}$, namely
{\fontsize{10pt}{11pt}\selectfont
\begin{align}
   &A\xrightarrow{\, {\mathbb Z}_{8}'\,}e^{\frac{2\pi i}{8}}A;\,\, \nn\\&
   A\xrightarrow{ {\mathbb Z}_{16}\times  {\mathbb Z}_{10}}e^{(9\cdot 2\frac{2\pi i}{16} +2\cdot 5\frac{2\pi i}{10}) }A =e^{\frac{2\pi i}{8}}A\,. \nn\\ 
   & \bar{F}\xrightarrow{\, {\mathbb Z}_{8}'\,}\bar{F};\nn\\&  \bar{F}\xrightarrow{ {\mathbb Z}_{16}\times  {\mathbb Z}_{10}}e^{(4\cdot 2\frac{2\pi i}{16} + 5\frac{2\pi i}{10})}\bar{F} =\bar{F}\,.
   \label{Eq:Z8ptoZ16} 
\end{align}}

Thus, we conclude that the largest anomaly-free discrete symmetry of the $SU(10)$ model, under which gauge-invariant operators carry non-zero charges is $\Gamma = {\mathbb Z}_{16}$. 
This implies that ${\mathbb Z}_{16}$ and its subgroups are the only possible discrete gauge symmetries that can be consistently gauged to accidentally generate the PQ symmetry. Their charges are listed in \cref{Tab:Z16andSubgroups}.
\begin{table}[h]
\begin{align*}
\begin{array}{|c|c|c|c|c|} 
\hline
& {\mathbb Z}_{16} & {\mathbb Z}_8 & {\mathbb Z}_4 & {\mathbb Z}_2 \\
\hline A & +9 & +1 & +1 & +1 \\
\hline \bar{F} & +4 & +4 & 0 & 0 \\
\hline\hline A \bar{F} \bar{F} & +1 & +1 & +1 & +1 \\
\hline {\rm Pf } A & +13 & +5 & +1 & +1\\
\hline
\end{array}
\end{align*}
\caption{Charge assignments of the anomaly-free discrete group ${\mathbb Z}_{16}$ and its subgroups.\label{Tab:Z16andSubgroups}}
\end{table}

\section{${\mathbb Z}_{16}$ gravitational anomalies}
\label{App:Z16anomalies}

As established in \cref{App:Discrete}, a ${\mathbb Z}_{16}$ subgroup of the $U(1)_{\rm PQ}$ symmetry remains anomaly free under QCD.  Furthermore, this ${\mathbb Z}_{16}$ subgroup does not have a ${\mathbb Z}_{16} \times [SU(10)]^2$ anomaly because it is a subgroup of $U(1)_{\rm PQ}$, which is free of $SU(10)$ anomalies. Similarly, by appropriately assigning hypercharges such that no fermion possesses a fractional charge, it can also be verified that the ${\mathbb Z}_{16} \times [U(1)_Y]^2$ anomaly is absent. Consequently, the ${\mathbb Z}_{16}$ discrete symmetry (or any of its subgroups) is anomaly free and can potentially be promoted to a $U(1)$ gauge symmetry.


In order to achieve this, it is necessary to check the gravitational anomalies which give rise to linear conditions on the charges~\footnote{Note that if the ${\mathbb Z}_{16}$ discrete symmetry is promoted to a $U(1)$ gauge symmetry, there is also a $U(1)^3$ anomaly condition. However, as discussed in Ref~\cite{Banks:1991xj,Csaki:1997aw}, this cubic anomaly condition on the U(1) charges is ambiguous and need not be satisfied because it depends on whether the high-energy theory above the $U(1)\rightarrow {\mathbb Z}_{16}$ breaking scale contains fractional charges.}; however, these can always be canceled by introducing light spectator fermionic singlets, which do not alter the dynamics of the model.
The gravitational anomalies of the ${\mathbb Z}_{16}$ symmetry and the subgroups indicated in \cref{Tab:Z16andSubgroups} are,
\begin{align}
\label{Eq: Z6 anomalies}
&{\mathbb Z}_{16}\times \left[grav\right]^2:\\  
&\qquad 45\cdot (+9)+60\cdot(+4)=5 \, ({\rm mod}\,  8)\nn\\ 
&{\mathbb Z}_{8}\times \left[grav\right]^2:\\  
&\qquad 45\cdot (+1)+60\cdot(+4)=5 = 1\, ({\rm mod}\, 4)\nn\\ 
&{\mathbb Z}_{4}\times \left[grav\right]^2:\\  
&\qquad 45\cdot (+1)=1 \, ({\rm mod }\,  2)\nn 
\end{align}
Note that due to the fact there are always even numbers of zero modes for a Weyl fermion in a gravitational instanton background, the ${\mathbb Z}_{N}\times \left[grav\right]^2$ anomaly has to be matched only mod $N/2$. For this reason we skip the ${\mathbb Z}_2$ symmetry since it always lacks a gravitational anomaly. In principle, the spectator fermions may allow for lower dimensional PQ-violating operators. However, as long as these fields do not acquire a VEV, they will not induce a displacement in the QCD axion potential minimum.

\newpage

\bibliographystyle{utphys}
\bibliography{BiblioNew}

\end{document}